\documentclass[10pt]{article}
\usepackage{graphicx}
\usepackage{fullpage}
\usepackage{subfigure}
\usepackage{url}
\usepackage{amsmath}
\usepackage{amsfonts}
\newcommand{\eat}[1]{}
\usepackage{overpic}

\usepackage[T1]{fontenc}
\usepackage[utf8]{inputenc}

\newcommand{\aba}{graph builder actor}
\newcommand{\sa}{sampler actor}

\newcommand{\rmc}{rigid molecular component}
\newcommand{\ctwo}{\ref{eqn:preferredConstraints}}
\newcommand{\cone}{\ref{eqn:constraints}}

\newcommand{\psone}{A}
\newcommand{\pstwo}{B}
\newcommand{\pone}{a}
\newcommand{\ptwo}{b}

\newcommand{\distall}{$C_1$}
\newcommand{\distexist}{$C_2$}

\newcommand{\dst}{\delta}  % lennard jones dist
\newcommand{\dlo}{\underline{\dst}} 
\newcommand{\dhi}{\overline{\dst}} 
\newcommand{\ijx}{{\pone\ptwo}}
\title{Parallel Online Directed Acyclic Graph Exploration for Atlasing Soft-Matter Assembly Configuration Spaces}
\author{ Rahul Prabhu \and Amit Verma \and Meera Sitharam }

\begin{document}
\maketitle

\begin{abstract}
The paper formalizes a version of parallel online directed acyclic graph (DAG) exploration, general enough to be readily 
mapped to many computational scenarios. In both the offline and online versions, vertices are weighted with the work units 
required for their processing, at least one parent must be completely processed before a child is processed, and at any 
given time only one processor can work on any given vertex. The online version has the following additional natural 
restriction:  only after a vertex is processed, are its required work units or its children known.
%the entire processing of a vertex is 
%assigned only to one processor, although it can be interrupted and resumed.

Using the Actor Model of parallel computation, it is shown that a natural class of parallel online algorithms meets 
a simple competitive ratio bound. We demonstrate and focus on the problem's occurrence in the scenario of energy 
landscape roadmapping or atlasing under pair-potentials, a highly compute-and-storage intensive modeling component 
integral to diverse applications involving soft-matter assembly. The method is experimentally validated using a C++ 
Actor Framework (CAF) software implementation built atop EASAL (Efficient Atlasing and Search of Assembly Landscapes), 
a substantial opensource software suite, running on multiple CPU cores of the HiperGator supercomputer, 
demonstrating linear speedup results.
\end{abstract}

\section{Introduction}
Energy landscape analysis in soft-matter assembly under pair-potentials is highly computationally 
intensive and occurs frequently in diverse application scenarios. The applications range from chemical 
engineering, material science, and computational chemistry to the study of microscale and nanoscale 
processes underlying life and disease \cite{Beltran-Villegas2011, 10.1371/journal.pcbi.1000415, Dock_rev1, Dock_rev2,kaku, Andricioaei_Karplus_2001, Killian_Yundenfreund_Kravitz_Gilson_2007, Wales2010, Wales-Sticky2018, Calvo2012}.
 
One integral component of the analysis, e.g., for equilibrium free energy and kinetics computations, 
is the task of roadmapping or atlasing  the  assembly configuration space into contiguous equi-potential 
energy regions. While this component is an  implicit requirement in all prevailing methodologies and is 
highly compute and storage intensive, only certain more recent methodologies  
\cite{canny-alg, bib:canny-roadmap, PrabhuEtAl2020, Ozkan2014MC, WuEtAl2020, PhysRevE, Wales:Landscapes, Holmes-Cerfon-2018, BaryshnikovEtAl2012, Baryshnikov08022013}, 
isolate it as an explicit component, using various strategies based on modern discrete geometry and algorithmic techniques 
to significantly enhance efficiency. 

EASAL (efficient atlasing and search of assembly landscapes) is a substantial opensource software suite \cite{easalVideo, easalSoftware, easalUserGuide} based on one such methodology \cite{Ozkan:toms, PrabhuEtAl2020} developed by  the authors, particularly suited to so-called short-ranged pair potentials that challenge prevailing methodologies.  See Section \ref{sec:easal_background}. EASAL treats the rigid components of assembling entities (molecules or particles) as point-sets and the  pair-potentials as distance interval constraints between each pair of points taken from different point-sets. 

Each contiguous equi-potential energy region  consists of configurations satisfies exactly a specific set of active pair-potential  constraints which uniquely specifies the region. The regions are topologically organized as nodes of a  directed acyclic graph (DAG), which is  a lattice poset -- of the corresponding sets of active pair-potential constraints -- under set containment.   

In Section \ref{sec:easal_background}, we demonstrate how the DAG is discovered or revealed gradually by sampling the configurations in each node or region starting from the root nodes which have no incoming edges.   Sampling a node or region reveals the immediate descendants (children) of the node, which can be sampled only thereafter. Sampling of a node is  a sequential process and the required effort is typically unknown apriori.

In Section \ref{sec:problem_setup}, we formalize a generalized version of the above atlasing problem  as a  parallel online  DAG exploration
starting from one or more  root nodes, with the following condition that only after a vertex is processed, are its children known (processing takes a nontrivial amount of computation that is unknown until processing is complete). We will see that it  follows that the entire processing of a vertex is 
assigned only to one processor in one contiguous block of work which does not need to be interrupted and resumed. 
%(i) discovering neighbors at any vertex requires a non-trivial amount of computation assigned as a weight of the vertex in the DAG  (ii) only after a vertex is processed,  are all its neighbors discovered, and only then can they be processed (iii)  the processing of a vertex  can be assigned only to one processor at any given time, although it can be interrupted and resumed.      

In fact,  many other computational problems far removed from assembly energy landscape atlasing, e.g., mapping of certain types of terrains using  drones or robots \cite{ota2006multi}
can be reduced to  this version of parallel online  DAG exploration.  

In Section \ref{sec:related_work}, we note that while there is extensive literature on parallel offline graph exploration \cite{exploring_unknown_graphs}, as well as a competitive ratio analysis for sequential online graph exploration \cite{exploring_unknown_graphs, exploring_unknown_graphs_efficiently},  to our knowledge, competitive ratio analysis for parallel online graph exploration is \cite{pajak2014algorithms} (see also Section \ref{sec:related_work}) is relatively limited, although there was substantial early work on the parallel online setting \cite{Aspnes1998}.

We give a lower bound on the competitive ratio \cite{Aspnes1998} of our version of parallel online graph exploration, i.e., the best performance of any parallel online algorithm in relation to the performance of the optimal parallel offline strategy.    
Tightness of this competitive ratio bound is demonstrated by showing that it is achieved by any online algorithm in a natural class that minimizes processor idle times by assigning available tasks immediately to processors about to become idle. 

In Section \ref{sec:algo}, we  formulate this class of parallel online algorithms using the \emph{actor model} of parallel computation and communication  for concurrent programming, in which the basic unit of computation is an abstract entity called an \emph{actor} 
(the actor model is discussed in more detail in Section \ref{sec:algo}). 

\begin{figure*}[htpb]
\centering
%Fig1
\subfigure[]
{\label{fig:exampleStratification} 
\begin{overpic}[scale=.2]{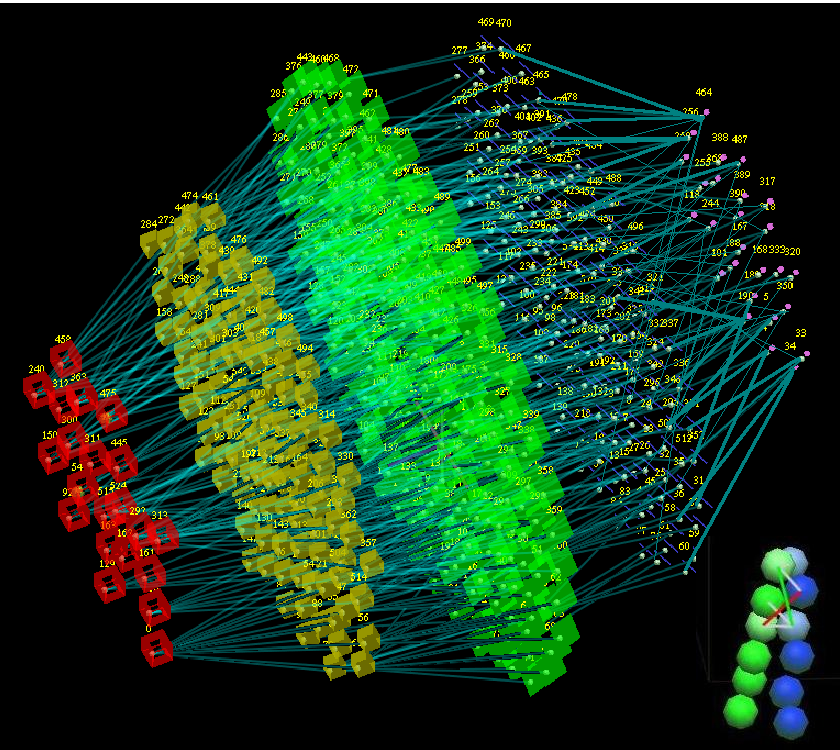}
\end{overpic}
}
%Fig2
\subfigure[]{
\begin{overpic}[scale=.135]{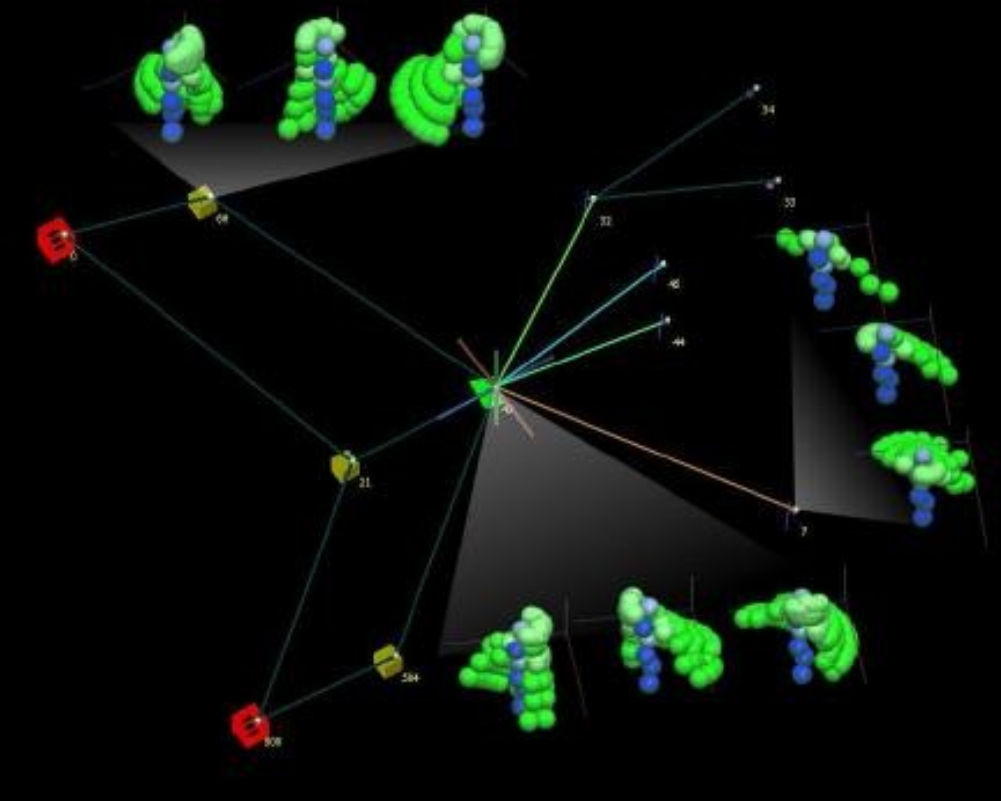}
\end{overpic}
\label{fig:pctreeSweep}
}
%Fig3
\subfigure[]{
\begin{overpic}[scale=.207]{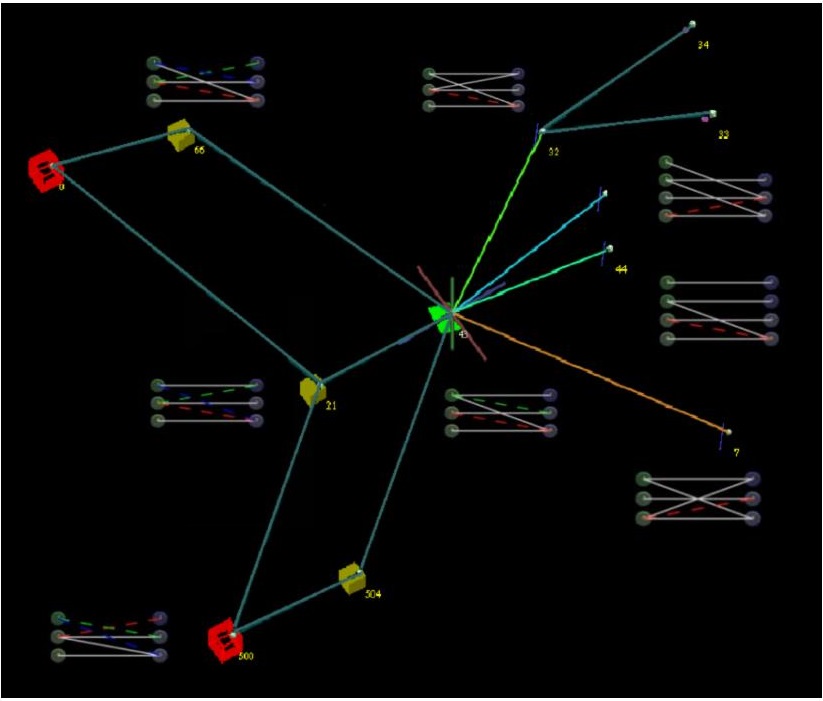}
\end{overpic}
\label{fig:pctreeACG}
}
\caption{\textbf{Stratification of the Roadmap}:
(a) shows a portion of the roadmap for the inset pair of input \rmc s. 
Each node represents a region of configurations satisfying a specific set of active constraints. 
The red nodes represent regions with 2 active 
constraints. Each successive stratum (from left to right) contains regions
with one additional active constraint until we reach 
the pink leaf nodes.
(b)  Regions in the roadmap shown with their Cartesian 
configuration sweeps. Each sweep is the union of Cartesian configurations 
in the corresponding  region. 
(c) Ancestors and descendants of a  node, shown 
with their active constraint sets. Figure adapted, courtesy \cite{PrabhuEtAl2020}.
}
\label{fig:NestedRegions}
\end{figure*}

In Section \ref{sec:results:verifying_speedup}, empirical results are presented to demonstrate a 
proof of concept for the effectiveness of our parallel scheme at achieving linear speedup  
for parallel exploration of assembly landscapes using the EASAL methodology. The parallel 
implementation is available as part of the EASAL opensource software suite \cite{easalSoftware}.
We use the C++ Actor Framework (CAF) \cite{chs-rapc-16, cshw-nassp-13} for the implementation.
The parallel implementation, running on the University of Florida's Hipergator supercomputer, on 
Intel(R) Xeon (TM) Gold 6142 series processors from the Skylake family series of processors give 
near linear speedup in the number of compute cores (as compared to the sequential implementation 
of the EASAL methodology \cite{easalSoftware}) and can roadmap several millions nodes in minutes, 
where the sequential version took hours.

A summary of contributions is listed in the abstract.

\subsection{Related Work on Parallel Offline, and  Sequential Online Graph Exploration}
\label{sec:related_work}
Graph traversal is the process of visiting all vertices in a connected component  of a graph (directed or undirected),  by starting from a chosen vertex and following edges. In the \emph{offline} 
version of the problem, the entire graph is available to the algorithm before the execution begins. The most common algorithms
for these are depth-first and breadth-first traversals, for which parallel
versions have been extensively studied \cite{naumov2017parallel, merrill2012scalable, chhugani2012fast, pearce2010multithreaded}.

In the \emph{online} version of graph traversal, often called graph exploration, the graph is discovered in stages during execution. Specifically, the (unexplored) neighborhood of a vertex is only visible after the vertex has been visited.
The basic version of the problem requires that the exploration algorithm visit every node and edge of the graph 
(directed or undirected) at least once. Doing this efficiently requires the algorithm to minimize the number of repeated 
edge traversals. Variants of the problem include requirements such as starting from a particular \emph{source} node and 
possibly returning to the source at the end. Commonly seen in robotics, it is a well-studied problem \cite{rao1993robot, das2019graph}. 
Online graph exploration algorithms, like all online algorithms, are typically evaluated by their competitive ratio,   defined as the ratio of the performance (according to some measure) of  the online algorithm (in the worst case) to that of the best offline strategy. 

There exist numerous sequential, online graph exploration methods that give different competitive ratios depending
on the class of input graphs.
The problem of exploring undirected graphs is solved using a DFS with a competitive ratio of 2 \cite{exploring_unknown_graphs}; while
 directed, strongly connected graphs yield a
 competitive ratio that is a factor of the \emph{deficiency} of the graph, which is the 
number of edges needed to make the graph Eulerian. In \cite{exploring_unknown_graphs} it is shown that the competitive
ratio is not bounded when the deficiency of the graph is unbounded and gives an online algorithm in which the 
complexity is bounded when the deficiency of the graph is bounded. 
Using a  simple depth-first strategy, \cite{kwek1997simple} gives an algorithm that is polynomial in 
the deficiency of the graph times the number of edges, for dense graphs. For the same problem, 
\cite{exploring_unknown_graphs_efficiently} gives an online algorithm whose competitive ratio is polynomial 
in the deficiency of the graph. 

\cite{panaite1999exploring}  gives an exploration algorithm whose \emph{penalty}, i.e., 
the worst-case number of edge traversals more than the lower bound (which is equal to the number of 
edges when each edge is visited only once),
of the order of the number of vertices in the graph. \cite{palacios2017random} gives a randomized
algorithm to improve efficiency and \cite{Bourdonov2004} gives an algorithm for traversing an 
unknown directed graph, based on the construction of the output-directed spanning tree of the graph and the 
breadth-first search on this tree.

 The related problem of online layered graph traversal \cite{fiat1998competitive}, which involves searching for a 
target vertex in a layered graph whose width and number of layers are unknown, has a tight lower bound on competitive ratio by \cite{RAMESH1995480}  for deterministic and randomized algorithms.

While parallel online algorithms overall were studied early on \cite{Aspnes1998}, the literature on \emph{parallel, online} graph  exploration is sparse, with the exception of  \cite{pajak2014algorithms, DERENIOWSKI201537, DERENIOWSKI2016802, CZYZOWICZ201770, MENC201717, KOSOWSKI201980, ivaskovic2017multiple}, 
who studies parallel online graph exploration using multiple agents. They give efficient algorithms 
and lower bounds for competitive ratios, for several families of graphs such as trees, cycles, random graphs, 
and cliques. To the best of our knowledge, this paper's version of parallel, online, vertex-weighted DAG exploration has not been studied.

\section{Problem Formalization and Competitive Ratio Analysis}
\label{sec:problem_setup}
We first define the \emph{offline} version of the problem. The input is a vertex-weighted DAG $G = (V,E,w: V\rightarrow \mathbb{N})$, where the sink vertices (no outgoing edges) $v$ have  $w(v) = 0$. Intuitively the vertex weight $w(v)$ denotes the amount of single processor time units it takes to  process $v$. In the \emph{sequential offline} version, the desired output is a traversal sequence $\{v_i\}$  satisfying the following property: For all  $v_l$ that are not sources, there is  an edge  $(v_j,v_l)\in E$  s.t.  $j \le l$, i.e.,  at least one parent  is  processed before a child.   

In the \emph{parallel offline} version with $r$ processors, the desired output is a \emph{work table}, namely a  function $S_v: {\cal T} = \{1,\ldots,T\} \rightarrow \{1,\ldots,r, *\}$, for each vertex $v$, assigning each time unit to at most one of the $r$ processors, or to $*$ denoting vertex idle time. The assignments $S_v$ must satisfy the following. (1) $|\{S^{-1}_v(p):1\le p \le r\}| = w(v)$ (total work done by processors on the vertex adds up to $w(v)$),  (2) For all $v$ that are not sources, there is some $(u,v) \in E$ s.t. $max \{t: S_u(t) \ne *\} \le min \{t: S_v(t) \ne *\}$  (at least one parent $u$ is processed completely before starting on a child $v$) and (3) for any time $t$, $u\ne v \Rightarrow S_u(t) \ne S_v(t)$, unless $S_u(t)=S_v(t) = *$, i.e.,  at any given time, a processor is assigned to  at most one vertex. The performance of the algorithm is given by $T$, the \emph{completion time}. Note that minimizing $T$  is equivalent to minimizing \emph{total processor idle time}, i.e., 
$\sum_{p} |\{t \in {\cal T}: \forall\  v S_v(t) \ne p \}|$. 

In the  \emph{online} version of the problem, not only is the DAG $G$ unknown apriori,  both the work $w(v)$ and the children of $v$ are unknown until $v$ is completely processed, i.e., until the time $max_t \{t: S_v(t) \ne *\}$.  Therefore, in the \emph{parallel, online} version without loss of generality, we can assume that each vertex is assigned to a unique processor which completes the work on it for a contiguous block of time. Formally, the parallel online version  adds an additional  \emph{simplified work table assumption}: for any work table $S_v$, there are times $1\le t_1,t_2 \le T$ s.t. $S_v(t)= *$ for $t< t_1$ and $t>t_2$,  and 
$S_v(t) = p$, for some processor $p$, $1\le p \le r$, when $t_1\le t\le t_2.$ It follows from the previously stated conditions on the offline version, that $w(v)= t_2-t_1+1$.

\noindent{\bf Fact.} When the number of processors $r=1$,  the \emph{sequential, online or offline} completion time $T$ reduces to  $\sum\limits_v w(v)$; and when $r \ge |V|$, the \emph{massively parallel, online or offline} completion time  $T = max_v\{w(v)\}$.

\begin{figure}
    \centering
    \subfigure[$r=2$]{\label{fig:2-processors}\includegraphics[width=0.33\linewidth]{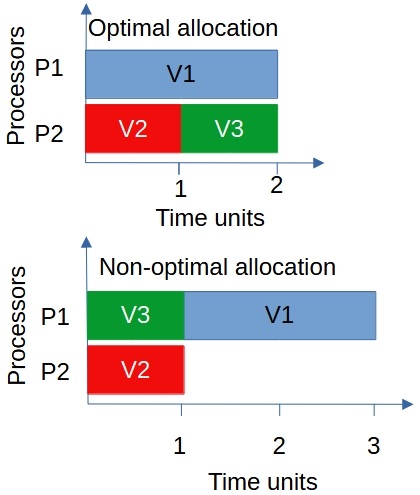}}
    \subfigure[$r=4$]{\label{fig:4-processors}\includegraphics[width=0.64\linewidth]{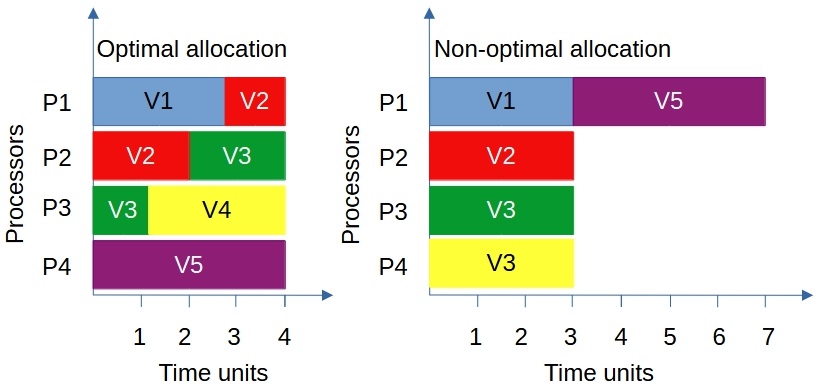}}
    \caption{Optimal and non-optimal allocation of tasks to processors. The horizontal axis 
    represents time units and the vertical axis shows the processors. 
    (a) In the optimal case, $T=2$, and in the non-optimal case, the $T=3$, making the competitive ratio 1.5. 
    (b) In the optimal case, $T=4$, and in the non-optimal case, $T=7$, giving a competitive ratio of 1.75.}
    \label{fig:processor-task-allocation}
\end{figure}
In addition to the above simplified work table assumption (wlog)  that minimizes movement of processors between vertices, we  define \emph{StayBusy}, a natural class of parallel, online, DAG exploration algorithms as satisfying the following additional, natural requirement.  The set of  idle vertices at any given time is globally communicated and is nonempty only if  the set of idle processors at that time is empty. Formally, at any given time $t$, the set $\{v: S_v(t) = *\} \ne \emptyset$, only if  $\{p: \not\exists\  v \ S^{-1}_v(p) = t\}  =  \emptyset$.
We state and sketch the proof of the following theorem.

\smallskip\noindent
{\bf Theorem.}
(1) The competitive ratio of the parallel online (vertex-weighted) DAG exploration problem with $r$ processors, defined as the ratio of the completion time (in the worst case) of the best online algorithm to that of the best offline strategy, is $2 - 1/r$.  (2) Every \emph{StayBusy} parallel online algorithm achieves this competitive ratio.

\noindent
\emph{Proof (Sketch)}:
The proof follows from the series of statements below.

\noindent
{\bf Proposition.}
 The optimal  parallel offline algorithm for the DAG traversal problem with $r$ processors  has completion time $T= max \{max_v\{W(v)\}, \  1/r\sum_v w(v)\}$; here $W(v) = \sum_{u\in \pi_v} w(u)$, where $\pi_v$ is the shortest path from any source vertex (no incoming edges) to $v$. 

   To prove the proposition, we first notice that even offline traversal requires at least one parent   to be processed before a child. Thus the  time to finish processing any sink vertex $v$ is at least  $W(v)$.  Since only one processor is assigned to any vertex at a given time, the total work $\sum_v w(v)$ cannot be completed by $r$ processors in  time less than $1/r \sum_v w(v)$.  Intuitively, it is also clear that this completion time is achieved by any parallel offline algorithm  that pre-computes a tree  of shortest paths $\pi_v$, and, at each time step $t$, assigns processors to the longest paths originating at uncompleted vertices with a completed parent.

 \noindent
{\bf Observation.}
Any \emph{StayBusy} algorithm for parallel online vertex-weighted DAG exploration is oblivious, i.e., for any isomorphism class of  vertex-weighted DAGs, all \emph{StayBusy} algorithms have the same completion time in the worst case. 
In particular, it is immaterial which data structure  (queue, stack) is used  at any given time $t$, to store and assign the idle vertex set $\{v: S_v(t) = *\}$ to idle processors $\{p: \not\exists\  v \ S^{-1}_v(p) = t\}$.
 
Here it is important to note that our analysis is based on deterministic algorithms and worst-case inputs, rather than randomized algorithms or average-case inputs. The next Lemma completes the proof of the Theorem.

\noindent
{\bf Lemma.}
The input DAG $G$ with the  worst-case competitive ratio for $r$ processors consists of just $r+1$  source vertices. For exactly one of the sources $v$, $w(v) = t^*$, and for the remaining sources $u$, $w(u) = (r-1)t^*/r$. For this input DAG $G$, the best parallel offline algorithm's completion time is $t^*$, and the best \emph{StayBusy} parallel online algorithm's completion time is $t^*(1+ (r-1)/r)$, yielding the worst-case competitive ratio of $2- 1/r$ in the theorem.  

A key intuition in the proof of the Lemma is that for a given number of processors $r$, the worst-case competitive ratio, i.e., the maximum advantage of the offline vs. the online algorithm occurs for inputs for which  the two quantities (whose maximum gives the completion time in the Proposition) are in fact equal, i.e., 
$T= max_v\{W(v)\} = 1/r\sum_v w(v)$.  For such inputs, the offline algorithm's advantage is maximized against the obliviousness of a \emph{StayBusy} algorithm, causing the highest total processor idle time. The worst-case input described in the Lemma satisfies this condition.
The proof has 3 parts. 
The first shows, in a series of steps,  how to decompose any input vertex-weighted DAG into a  sequence of inputs as in the Lemma, while  ensuring the competitive ratio of  the best \emph{Staybusy} algorithm remains no worse than the $2-1/r$  in the theorem. The first step shows that  for the same type of input DAG, if the number of sources $m$ exceeds $r+1$, or if the weights of the sources are unequal, the competitive ratio only improves. For example, in the former case, the competitive ratio is $\lfloor m-1/r\rfloor *(r-1)/(m-1) + 1$, which is less than $2- 1/r = (r-1)/r+1$ when $m > r+1.$ The second step shows that input DAGs that are the union of disjoint paths can be reduced to the above simple case. The third step decomposes DAGs that are constructed starting from a union of disjoint paths (each starting from a source vertex), with additional edges that cross between the paths, and the final step  adds additional paths that originate from non-source vertices: in both cases we show how to decompose the DAG into a sequence of simple worst-case inputs as in the Lemma, whose combined competitive ratio is preserved. 
   
The remaining two parts of the proof of the Lemma are straightforward and demonstrate the completion times of the best parallel offline and online algorithms with $r$ processors, and thus the competitive ratio, for this worst-case input. See Figure \ref{fig:processor-task-allocation}.

\section{Parallel Online DAG Exploration in the Actor Model}
\label{sec:algo}
In this section, we describe our parallel online DAG traversal algorithm, easily seen to belong to  the \emph{StayBusy} class using the \emph{actor model,}
a concurrent programming model in which the basic unit of computation is an abstract entity called an \emph{actor}. 
Actors communicate asynchronously through message passing and have well-defined behaviors for each
message they receive. In response to messages, actors can make local decisions, reply to messages, 
or spawn more actors. Actors don't share state, avoiding critical sections and race conditions commonly 
plaguing concurrent programs. This prevents the need for synchronization through locks, leading to less
complicated code and better performance. 

Our algorithm for online DAG traversal uses two types of actors, a \emph{graphbuilder actor} and
multiple worker actors which we call \emph{sampler actors} in the assembly landscape atlasing context. See Section \ref{sec:easal_background} for the mapping of the formal problem in Section \ref{sec:problem_setup} to the assembly landscape atlasing scenario. The graphbuilder actor keeps track of the graph 
discovered so far in its local state and acts as a central point of communication between the sampler actors. 
The sampler actors process or sample the nodes of the graph to discover their neighbors. Upon discovery, they 
communicate the unique address of the nodes to the graphbuilder actor. Communication of the newly discovered 
connections lets the graphbuilder actor spawn new actors (constrained by the available number of processors $r$) to sample the newly discovered nodes and 
prevents repeated processing of nodes discovered through multiple paths in the graph. 

%Figure \ref{fig:workflow} shows the overall workflow. 
The algorithm starts with the main function sending a Start message
to the graphbuilder actor, which the \aba\ initializes an empty graph
data structure to store the DAG explored so far, creates an unprocessed or unsampled nodes queue, 
and adds the source nodes to it. Next, it spawns as many sampler actors as the number of 
idle processors available and messages them the description of the source nodes for them to sample and explore. 
This, together with the unprocessed nodes queue discussed below ensure the \emph{StayBusy} condition is met.

A \sa\ starts when it receives the description of the node to be sampled as input from the \aba. 
During sampling, any new nodes found are sent to the \aba\ as a sampling result message, to 
be recorded in the DAG and to be independently, recursively explored in parallel. 

Upon receiving the sampling result message containing the description of neighbor nodes, 
the \aba\ checks the graph data structure to determine if the nodes  in
the message have been previously explored. The nodes  that haven't yet been explored
are added it to the unprocessed or unsampled nodes queue. Appropriate edge connections are
added in the graph data structure. This ensures the \emph{StayBusy} condition:  at any given 
time $t$, the set $\{v: S_v(t) = *\} \ne \emptyset$, only if  $\{p: \not\exists\  v \ S^{-1}_v(p) = t\}  =  \emptyset$.

\section{Empirical Results: Parallel Implementation of Assembly Landscape Atlasing with Linear Speedup}
\label{sec:results}
Section \ref{sec:easal_background} maps the online graph exploration problem of \ref{sec:algo} 
to the application context of molecular assembly configuration space atlasing. Section 
\ref{sec:results:verifying_speedup} presents empirical results demonstrating a proof of concept 
for the effectiveness of our parallel scheme at achieving linear speedup for
parallel exploration of assembly landscapes using the EASAL methodology.

\subsection{Online DAG exploration in Assembly Landscape Atlasing}
\label{sec:easal_background}
EASAL (efficient atlasing and search of assembly landscapes) \cite{Ozkan:toms, PrabhuEtAl2020} treats the rigid 
components of assembling entities (molecules or particles) as point-sets and the  pair-potentials as distance interval 
between each pair of points taken from different point-sets. The EASAL methodology takes as input a 
collection of $k$ point-sets with at most $n$ points, each specified as the positions of the $n$ points, and 
distance interval constraints on point pairs  $(a, b)$ given by $\dlo_\ijx, \dhi_\ijx \in \mathbb{R}_+$ (see Equations 
\cone\  and \ctwo, which are formulated for $k=2$). 
Equations \cone\ requires that for all point pairs, one from each point set, their distances are greater than $\dlo_\ijx$
and equation \ctwo\ requires that at least one point pair, one from each point set, be at a \emph{desired} distance 
$\dhi_\ijx$.
\begin{align}
\forall (\pone \in \psone, \ptwo \in \pstwo),\qquad& \dst_\ijx \ge \dlo_\ijx& \dlo_\ijx \in \mathbb{R}_+ \tag{\distall}\label{eqn:constraints}\\
\exists (\pone \in \psone, \ptwo \in \pstwo),\qquad& \dst_\ijx  \le \dhi_\ijx, & \dhi_\ijx \in \mathbb{R}_+.\tag{\distexist}\label{eqn:preferredConstraints}
\end{align}

The output of EASAL is the topological and geometric structure of the assembly landscape 
i.e., the set of all feasible $T \in SE(3)$ satisfying (\cone, \ctwo). For general $k$,  Equation \ctwo\ is generalized to the minimum number of constraints to ensure that the feasible configuration space is bounded. It is divided into
contiguous equi-potential energy regions   each of which  
 consists of configurations satisfying 
exactly a specific set of `active' constraints of type \ctwo\ which uniquely specifies
the region. The regions are topologically organized as nodes of a  DAG
$(V,E,w: V\rightarrow \mathbb{N})$ as in Section \ref{sec:problem_setup}, where each edge points 
from a parent set of active constraints to a child superset containing one additional active constraint. 
Thus the atlas  is a lattice poset -- of the corresponding sets of active constraints -- under set containment.  
 The DAG is discovered or revealed gradually by sampling the configurations in each node or region starting from the source nodes which have the minimal number of active constraints required to ensure that the configuration space is bounded.   Sampling a node or region reveals the immediate descendants (children) of the node, which can be sampled only thereafter. Sampling of a node $v$ is  a sequential process and the required effort $w(v)$ is typically unknown until the processing is complete.

\subsection{Speedup of the Parallel Implementation}
\label{sec:results:verifying_speedup}
The experiments were run on the Hipergator supercomputer with a varying number
of Intel(R) Skylate(TM) series processors all on the same switch, in a single
node of the cluster, to reduce latency in message passing. Suitable g++ compiler 
optimizations were used on the code to enhance its performance.

\begin{figure}[htpb] \centering
\subfigure[]{\label{fig:6Straight}\includegraphics[scale=0.08]{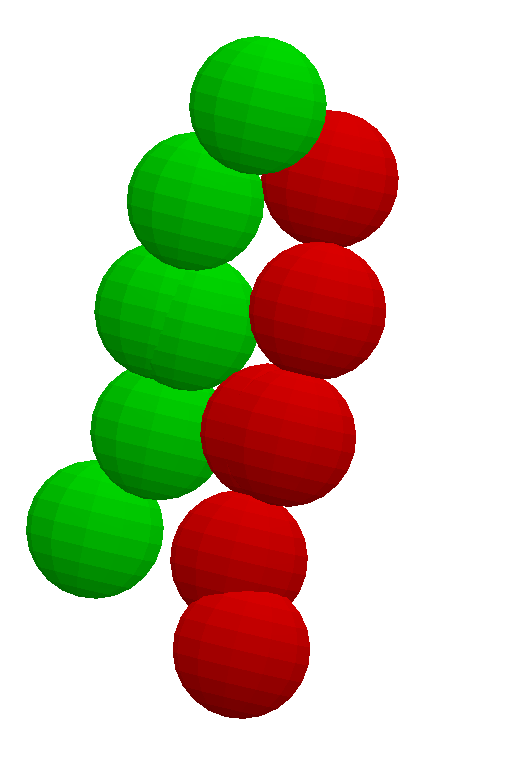}}
\subfigure[]{\label{fig:6Pocketed}\includegraphics[scale=0.08]{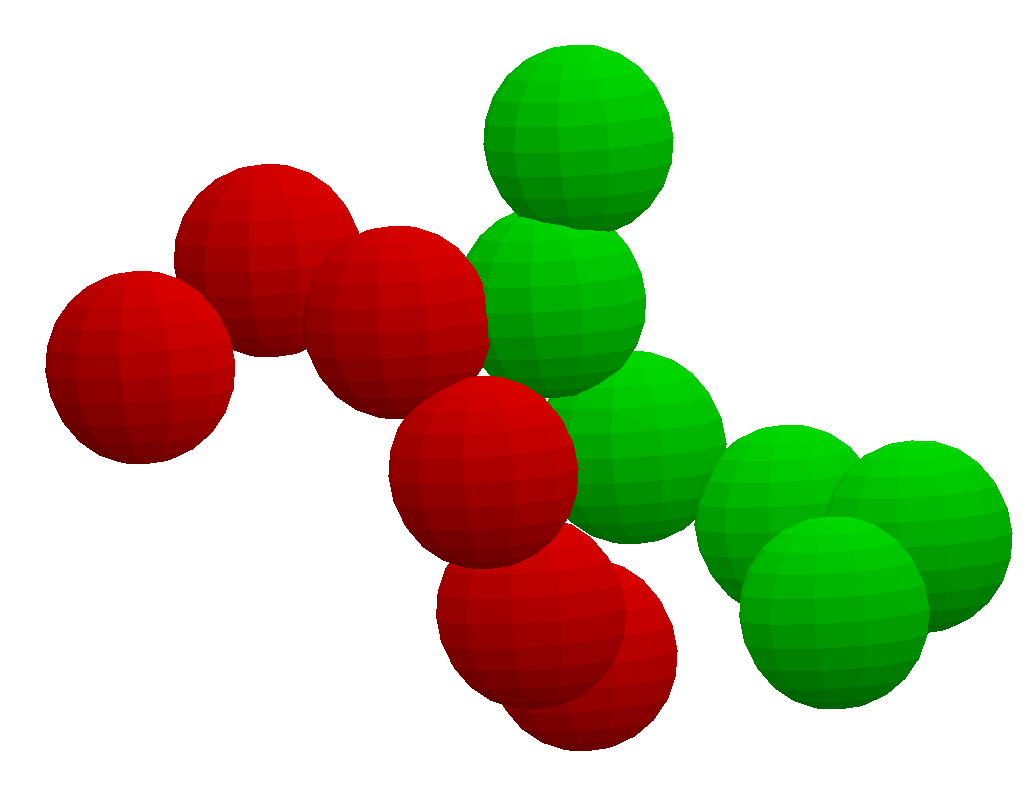}}
\subfigure[]{\label{fig:10Pocketed}\includegraphics[scale=0.08]{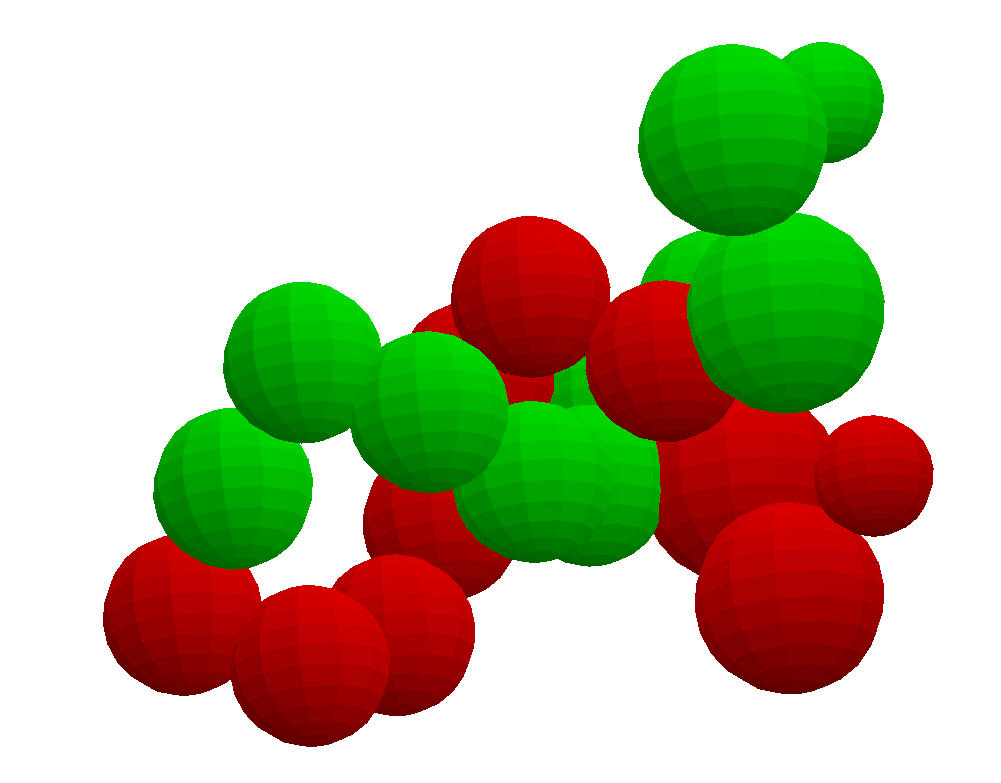}}
%\subfigure[]{\label{fig:20Pocketed}\includegraphics[scale=0.08]{figures/20Pocketed.png}}
\subfigure[]{\label{fig:20Straight}\includegraphics[scale=0.08]{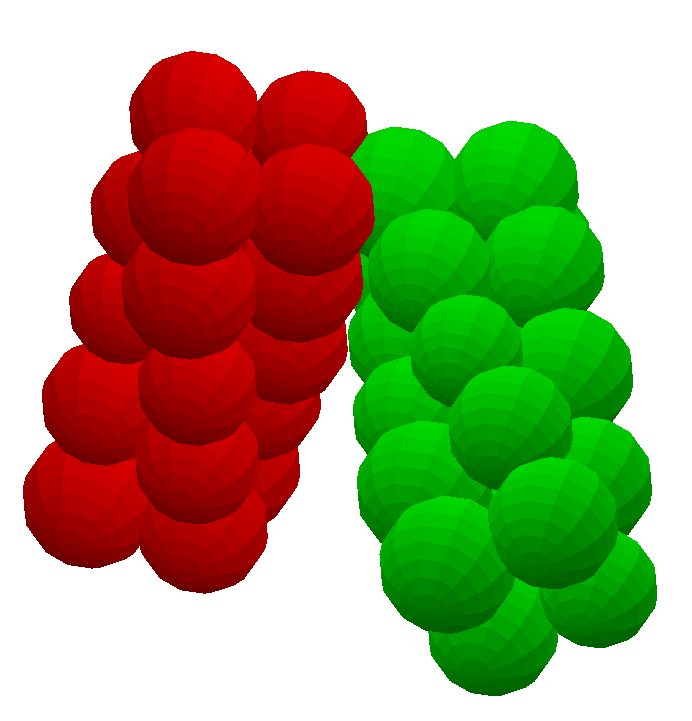}}
%\subfigure[]{\label{fig:42Pocketed}\includegraphics[scale=0.05]{figures/42Pocketed.png}}
\caption{\textbf{List of \rmc s used in the experiments.}: (a) Narrow Convex (6 Atoms). (b) Narrow Concave (6
Atoms). (c) Narrow Concave (10 Atoms) %(d) Wide Convex (20 Atoms). 
(d) Wide Concave (20 Atoms).} %(e) Wide Concave (42 Atoms).}
\label{fig:inputMolecules}
\end{figure}

The primary computation was atlas generation. We generate complete atlases for the 
input molecules described in Figure \ref{fig:inputMolecules}. 
The parallel implementation run on a single core was slightly faster than the 
original sequential implementation of \cite{PrabhuEtAl2020}, due to actors being able to 
take advantage of the idle times of other actors.
However,  with more processors, the speed-up is essentially linear even when 
compared to the original sequential implementation. To be able to compare the
results across \rmc s, we perform normalizations similar to  
the single threaded experiments in the paper \cite{PrabhuEtAl2020}. In particular,
we fix the ratio of the average atom radius to the sampling step size $t$. We
additionally fix the width of the Lennard-Jones' well, $\overline{\delta_{ab}}
- \delta{ab}$ for an atom pair $(a, b)$,
with radii $\rho_a$ and $\rho_b$, to $0.25 * (\rho_a + \rho_b)$. Each of
these input assembly systems was sampled with 3 different values of $t$, to
analyze the effects of step size on the sampling time. In addition, each input
was run 10 times and the results were averaged to account for variability in
Hipergator loads.
\begin{figure}[htpb]
    \centering
    \includegraphics[width=0.9\linewidth]{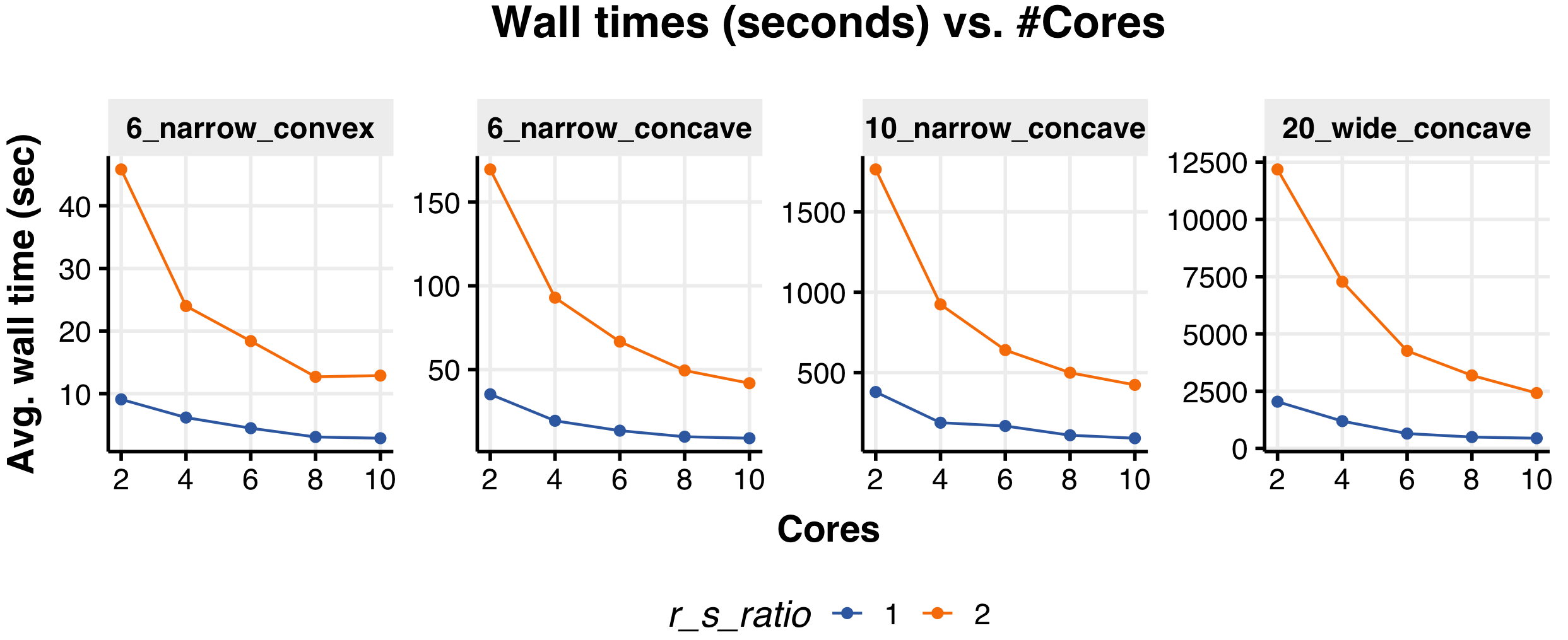}
    \caption{{The number of compute cores used on the x-axis and the 
    normalized wall time for atlasing the molecules (indicated at the top of every sub-figure) on the y-axis. 
    Here r\_s\_ratio is the ratio of the average atomic radius in the molecule to the stepsize.}}
    \label{fig:normalized_wall_time_steps}
\end{figure}

In Figure \ref{fig:normalized_wall_time_steps} 
we plot the normalized wall time for atlas sampling on the y-axis, and the number 
of compute cores used on the x-axis, for the same molecule but with different 
step sizes. The smaller the step size, the longer the sampling subroutine takes 
to sample each atlas node (see complexity analysis in \cite{PrabhuEtAl2020}). By 
parallelizing the sampling of atlas nodes, we see the highest benefit when the 
step size is the smallest.

\begin{figure}[htpb]
    \centering
    \includegraphics[width=0.9\linewidth]{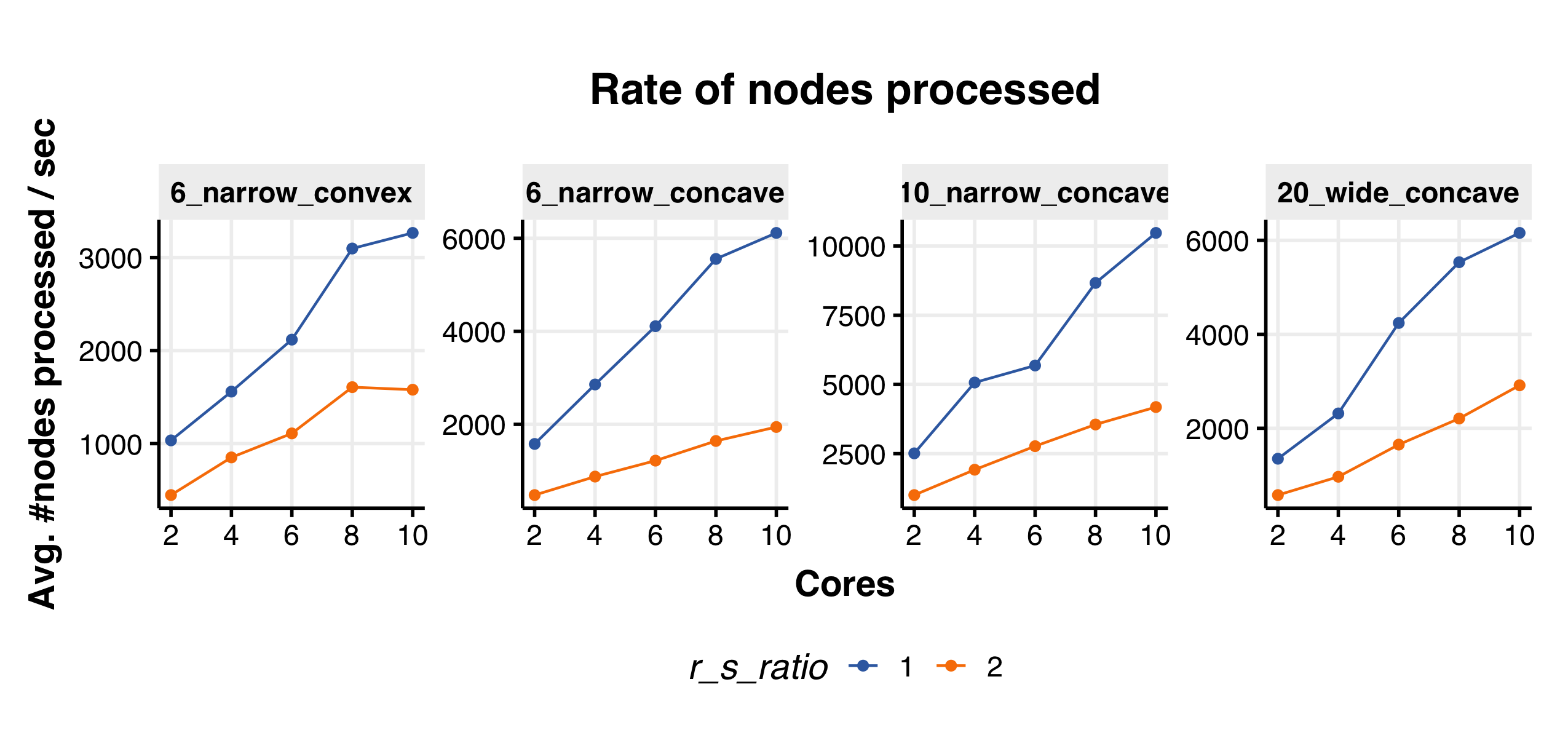}
    \caption{{The number of compute cores on the x-axis and the average number of atlas nodes discovered per
second on the y-axis, while atlasing the molecules (indicated at the top of every sub-figure). 
Here r\_s\_ratio is the ratio of the average atomic radius in the molecule to the stepsize.}} 
    \label{fig:rate_of_node_discovery}
\end{figure}

Figure \ref{fig:rate_of_node_discovery} plots the number of compute cores used on 
the x-axis and the number of graph nodes processed per second on the y-axis input molecules
indicated at the top of each sub-figure. As can be seen from
the figure, the rate of nodes processed increases linearly as the number of compute cores increases.

\section{Conclusion}
\label{sec:conclusion}
In this article, we formalize a version of parallel online directed acyclic graph (DAG) exploration with defined natural features and demonstrate the problem's occurrence in the highly compute-and-storage intensive modeling area of energy landscape roadmapping or atlasing under pair-potentials.
Using the Actor Model of computation, we show that a natural class of parallel online algorithms meets a simple competitive ratio bound.  The C++ Actor Framework (CAF) software implementation built atop  EASAL (Efficient Atlasing and Search of Assembly Landscapes),  running on multiple CPU cores of the HiperGator supercomputer, is used to demonstrate linear speedup results.

\subsubsection*{Acknowledgment}
%\textbf{Funding}
Authors Rahul Prabhu and Meera Sitharam were supported in part by NSF Grants DMS-1563234, and DMS-1564480. The authors acknowledge University of Florida Research Computing for providing computing resources. URL: http://www.rc.ufl.edu.

\bibliographystyle{plain}
\bibliography{references}

\begin{thebibliography}{10}

\bibitem{Andricioaei_Karplus_2001}
Ioan Andricioaei and Martin Karplus.
\newblock On the calculation of entropy from covariance matrices of the atomic fluctuations.
\newblock {\em The Journal of Chemical Physics}, 115(14):6289--6292, 2001.

\bibitem{Aspnes1998}
James Aspnes.
\newblock {\em Competitive analysis of distributed algorithms}, pages 118--146.
\newblock Springer Berlin Heidelberg, Berlin, Heidelberg, 1998.

\bibitem{Baryshnikov08022013}
Yuliy Baryshnikov, Peter Bubenik, and Matthew Kahle.
\newblock Min-type morse theory for configuration spaces of hard spheres.
\newblock {\em International Mathematics Research Notices}, 2014(9):2577--2592, 2014.

\bibitem{Beltran-Villegas2011}
Daniel~J. Beltran-Villegas and Michael~A. Bevan.
\newblock {Free energy landscapes for colloidal crystal assembly}.
\newblock {\em Soft Matter}, 7(7):3280--3285, 2011.

\bibitem{Bourdonov2004}
I.~B. Bourdonov.
\newblock Traversal of an unknown directed graph by a finite robot.
\newblock {\em Programming and Computer Software}, 30(4):188--203, Jul 2004.

\bibitem{Calvo2012}
Florent Calvo, Jonathan P~K Doye, and David~J Wales.
\newblock {Energy landscapes of colloidal clusters: thermodynamics and rearrangement mechanisms.}
\newblock {\em Nanoscale}, 4(4):1085--1100, 2012.

\bibitem{canny-alg}
J.~Canny.
\newblock A new algebraic method for robot motion planning and real geometry.
\newblock In {\em 28th Annual Symposium on Foundations of Computer Science (sfcs 1987)}, pages 39--48, 1987.

\bibitem{bib:canny-roadmap}
John Canny.
\newblock Computing roadmaps of general semi-algebraic sets.
\newblock {\em Computer Journal}, 36:504--514, 1993.

\bibitem{PhysRevE}
Gunnar Carlsson, Jackson Gorham, Matthew Kahle, and Jeremy Mason.
\newblock Computational topology for configuration spaces of hard disks.
\newblock {\em Phys. Rev. E}, 85:011303 1--8, 2012.

\bibitem{BaryshnikovEtAl2012}
Gunnar Carlsson, Jackson Gorham, Matthew Kahle, and Jeremy Mason.
\newblock Computational topology for configuration spaces of hard disks.
\newblock {\em Phys. Rev. E}, 85:011303, Jan 2012.

\bibitem{chs-rapc-16}
Dominik Charousset, Raphael Hiesgen, and Thomas~C. Schmidt.
\newblock {Revisiting Actor Programming in C++}.
\newblock {\em Computer Languages, Systems \& Structures}, 45:105--131, April 2016.

\bibitem{cshw-nassp-13}
Dominik Charousset, Thomas~C. Schmidt, Raphael Hiesgen, and Matthias W{\"a}hlisch.
\newblock {Native Actors -- A Scalable Software Platform for Distributed, Heterogeneous Environments}.
\newblock In {\em Proc. of the 4rd ACM SIGPLAN Conference on Systems, Programming, and Applications (SPLASH '13), Workshop AGERE!}, pages 87--96, New York, NY, USA, Oct. 2013. ACM.

\bibitem{chhugani2012fast}
Jatin Chhugani, Nadathur Satish, Changkyu Kim, Jason Sewall, and Pradeep Dubey.
\newblock Fast and efficient graph traversal algorithm for cpus: Maximizing single-node efficiency.
\newblock In {\em 2012 IEEE 26th International Parallel and Distributed Processing Symposium}, pages 378--389. IEEE, 2012.

\bibitem{CZYZOWICZ201770}
Jurek Czyzowicz, Dariusz Dereniowski, Leszek GÄ…sieniec, Ralf Klasing, Adrian Kosowski, and Dominik PajÄ…k.
\newblock Collision-free network exploration.
\newblock {\em Journal of Computer and System Sciences}, 86:70--81, 2017.

\bibitem{das2019graph}
Shantanu Das.
\newblock Graph explorations with mobile agents.
\newblock {\em Distributed Computing by Mobile Entities: Current Research in Moving and Computing}, pages 403--422, 2019.

\bibitem{exploring_unknown_graphs}
Xiaotie Deng and Christos~H. Papadimitriou.
\newblock Exploring an unknown graph.
\newblock {\em Journal of Graph Theory}, 32(3):265--297, 1999.

\bibitem{DERENIOWSKI201537}
Dariusz Dereniowski, Yann Disser, Adrian Kosowski, Dominik PajÄ…k, and PrzemysÅ‚aw UznaÅ„ski.
\newblock Fast collaborative graph exploration.
\newblock {\em Information and Computation}, 243:37--49, 2015.
\newblock 40th International Colloquium on Automata, Languages and Programming (ICALP 2013).

\bibitem{DERENIOWSKI2016802}
Dariusz Dereniowski, Adrian Kosowski, Dominik PajÄ…k, and PrzemysÅ‚aw UznaÅ„ski.
\newblock Bounds on the cover time of parallel rotor walks.
\newblock {\em Journal of Computer and System Sciences}, 82(5):802--816, 2016.

\bibitem{fiat1998competitive}
Amos Fiat, Dean~P Foster, Howard Karloff, Yuval Rabani, Yiftach Ravid, and Sundar Vishwanathan.
\newblock Competitive algorithms for layered graph traversal.
\newblock {\em SIAM Journal on Computing}, 28(2):447--462, 1998.

\bibitem{exploring_unknown_graphs_efficiently}
Rudolf Fleischer and Gerhard Trippen.
\newblock Exploring an unknown graph efficiently.
\newblock In Gerth~St{\o}lting Brodal and Stefano Leonardi, editors, {\em Algorithms -- ESA 2005}, pages 11--22, Berlin, Heidelberg, 2005. Springer Berlin Heidelberg.

\bibitem{ivaskovic2017multiple}
Andrej Ivaskovic, Adrian Kosowski, Dominik Pajak, and Thomas Sauerwald.
\newblock Multiple random walks on paths and grids.
\newblock In {\em 34th Symposium on Theoretical Aspects of Computer Science (STACS 2017)}. Schloss-Dagstuhl-Leibniz Zentrum f{\"u}r Informatik, 2017.

\bibitem{Dock_rev1}
Joel Janin.
\newblock Protein-protein docking tested in blind predictions: the capri experiment.
\newblock {\em Mol. BioSyst.}, 6(12):2351--2362, 2010.

\bibitem{kaku}
M.~Karplus and J.N. Kushick.
\newblock Method for estimating the configurational entropy of macromolecules.
\newblock {\em Macromolecules}, 14(2):325--332, 1981.

\bibitem{Killian_Yundenfreund_Kravitz_Gilson_2007}
Benjamin~J Killian, Joslyn Yundenfreund~Kravitz, and Michael~K Gilson.
\newblock Extraction of configurational entropy from molecular simulations via an expansion approximation.
\newblock {\em The Journal of chemical physics}, 127(2):024107--1 -- 024107--16, 2007.

\bibitem{KOSOWSKI201980}
Adrian Kosowski and Dominik PajÄ…k.
\newblock Does adding more agents make a difference? a case study of cover time for the rotor-router.
\newblock {\em Journal of Computer and System Sciences}, 106:80--93, 2019.

\bibitem{kwek1997simple}
Stephen Kwek.
\newblock On a simple depth-first search strategy for exploring unknown graphs.
\newblock In {\em Workshop on Algorithms and Data Structures}, pages 345--353. Springer, 1997.

\bibitem{MENC201717}
Artur Menc, Dominik PajÄ…k, and PrzemysÅ‚aw UznaÅ„ski.
\newblock Time and space optimality of rotor-router graph exploration.
\newblock {\em Information Processing Letters}, 127:17--20, 2017.

\bibitem{merrill2012scalable}
Duane Merrill, Michael Garland, and Andrew Grimshaw.
\newblock Scalable gpu graph traversal.
\newblock {\em ACM Sigplan Notices}, 47(8):117--128, 2012.

\bibitem{naumov2017parallel}
Maxim Naumov, Alysson Vrielink, and Michael Garland.
\newblock Parallel depth-first search for directed acyclic graphs.
\newblock In {\em Proceedings of the Seventh Workshop on Irregular Applications: Architectures and Algorithms}, pages 1--8, 2017.

\bibitem{ota2006multi}
Jun Ota.
\newblock Multi-agent robot systems as distributed autonomous systems.
\newblock {\em Advanced engineering informatics}, 20(1):59--70, 2006.

\bibitem{Ozkan:toms}
Aysegul Ozkan, Rahul Prabhu, Troy Baker, James Pence, Jorg Peters, and Meera Sitharam.
\newblock Algorithm 990: Efficient atlasing and search of configuration spaces of point-sets constrained by distance intervals.
\newblock {\em ACM Trans. Math. Softw.}, 44(4):48:1--48:30, 2018.

\bibitem{easalSoftware}
Aysegul Ozkan, Rahul Prabhu, Troy Baker, James Pence, and Meera Sitharam.
\newblock Efficient atlasing and search of assembly landscapes, 2021.
\newblock EASAL software. Available Online: https://bitbucket.org/geoplexity/easal.

\bibitem{Ozkan2014MC}
Aysegul Ozkan, Meera Sitharam, Jose~C. Flores-Canales, Rahul Prabhu, and Maria Kurnikova.
\newblock {Baseline Comparisons of Complementary Sampling Methods for Assembly Driven by Short-Ranged Pair-Potentials towards Fast and Flexible Hybridization}.
\newblock {\em Journal of Chemical Theory and Computation}, 2021.

\bibitem{pajak2014algorithms}
Dominik Pajak.
\newblock {\em Algorithms for deterministic parallel graph exploration}.
\newblock PhD thesis, Universit{\'e} Sciences et Technologies-Bordeaux I, 2014.

\bibitem{palacios2017random}
Alfredo~Toriz Palacios, Abraham S{\'a}nchez~L, and Jose Mar{\'\i}a~Enrique Bedolla~Cordero.
\newblock The random exploration graph for optimal exploration of unknown environments.
\newblock {\em International Journal of Advanced Robotic Systems}, 14(1):1729881416687110, 2017.

\bibitem{panaite1999exploring}
Petri{\c{s}}or Panaite and Andrzej Pelc.
\newblock Exploring unknown undirected graphs.
\newblock {\em Journal of Algorithms}, 33(2):281--295, 1999.

\bibitem{pearce2010multithreaded}
Roger Pearce, Maya Gokhale, and Nancy~M Amato.
\newblock Multithreaded asynchronous graph traversal for in-memory and semi-external memory.
\newblock In {\em SC'10: Proceedings of the 2010 ACM/IEEE International Conference for High Performance Computing, Networking, Storage and Analysis}, pages 1--11. IEEE, 2010.

\bibitem{easalVideo}
Rahul Prabhu, Troy Baker, and Meera Sitharam.
\newblock Video illustrating the opensource software {EASAL}, 2016.

\bibitem{easalUserGuide}
Rahul Prabhu and Meera Sitharam.
\newblock {EASAL} software user guide., 2016.

\bibitem{PrabhuEtAl2020}
Rahul Prabhu, Meera Sitharam, Aysegul Ozkan, and Ruijin Wu.
\newblock Atlasing of assembly landscapes using distance geometry and graph rigidity.
\newblock {\em Journal of Chemical Information and Modeling, to appear}, 2020.

\bibitem{10.1371/journal.pcbi.1000415}
Diego Prada-Gracia, Jes\'us G\'omez-Garde\=nes, Pablo Echenique, and Fernando Falo.
\newblock Exploring the free energy landscape: From dynamics to networks and back.
\newblock {\em PLoS Comput Biol}, 5(6):1--9, 2009.

\bibitem{RAMESH1995480}
H.~Ramesh.
\newblock On traversing layered graphs on-line.
\newblock {\em Journal of Algorithms}, 18(3):480--512, 1995.

\bibitem{rao1993robot}
N~SV Rao, Srikumar Kareti, Weimin Shi, and S~Sitharama Iyengar.
\newblock Robot navigation in unknown terrains: Introductory survey of non-heuristic algorithms.
\newblock Technical report, Oak Ridge National Lab.(ORNL), Oak Ridge, TN (United States), 1993.

\bibitem{Dock_rev2}
David~W. Ritchie.
\newblock Recent progress and future directions in protein-protein docking.
\newblock {\em Current Protein and Peptide Science}, 9:1--15, 2008.
\newblock doi:10.2174/138920308783565741.

\bibitem{Wales-Sticky2018}
Lukas Trombach, Robert~S. Hoy, David~J. Wales, and Peter Schwerdtfeger.
\newblock From sticky-hard-sphere to lennard-jones-type clusters.
\newblock {\em Phys. Rev. E}, 97:043309 1--10, 2018.

\bibitem{Wales2010}
David~J Wales.
\newblock {Energy landscapes of clusters bound by short-ranged potentials.}
\newblock {\em Chemphyschem : a European journal of chemical physics and physical chemistry}, 11(12):2491--2494, 2010.

\bibitem{Wales:Landscapes}
David~J. Wales.
\newblock Exploring energy landscapes.
\newblock {\em Annual Review of Physical Chemistry}, 69(1):401--425, 2018.
\newblock PMID: 29677468.

\bibitem{WuEtAl2020}
Ruijin Wu, Rahul Prabhu, Antonette Bennett, Aysegul Ozkan, Mavis Agbandje-McKenna, and Meera Sitharam.
\newblock {Prediction of Crucial Interaction for Icosahedral Capsid Self-Assembly by Configuration Space Atlasing using EASAL}, 2020.
\newblock Preprint available on arXiv: https://arxiv.org/abs/2001.00316.

\bibitem{Holmes-Cerfon-2018}
Emilio Zappa, Miranda Holmes-Cerfon, and Jonathan Goodman.
\newblock Monte carlo on manifolds: Sampling densities and integrating functions.
\newblock {\em Communications on Pure and Applied Mathematics}, 71(12):2609--2647, 2018.

\end{thebibliography}
\end{document}